\documentclass[12pt]{article}

\def\fnote#1#2{\begingroup\def\thefootnote{#1}\footnote{#2}\addtocounter{footnote}{-1}\endgroup}

\usepackage{amsmath}
\usepackage{amssymb}
\usepackage{makeidx}
\usepackage{epsf}
\usepackage{graphicx}        

\makeindex

\def\inbar{\vrule height1.5ex width.4pt depth0pt}
\def\IB{\relax{\rm I\kern-.18em B}}
\def\IC{\relax\,\hbox{$\inbar\kern-.3em{\rm C}$}}
\def\ID{\relax{\rm I\kern-.18em D}}
\def\IE{\relax{\rm I\kern-.18em E}}
\def\IF{\relax{\rm I\kern-.18em F}}
\def\IG{\relax\,\hbox{$\inbar\kern-.3em{\rm G}$}}
\def\IH{\relax{\rm I\kern-.18em H}}
\def\II{\relax{\rm I\kern-.18em I}}
\def\IK{\relax{\rm I\kern-.18em K}}
\def\IL{\relax{\rm I\kern-.18em L}}
\def\IM{\relax{\rm I\kern-.18em M}}
\def\IN{\relax{\rm I\kern-.18em N}}
\def\IO{\relax\,\hbox{$\inbar\kern-.3em{\rm O}$}}
\def\IP{\relax{\rm I\kern-.18em P}}
\def\IQ{\relax\,\hbox{$\inbar\kern-.3em{\rm Q}$}}
\def\IR{\relax{\rm I\kern-.18em R}}
\def\IT{\relax{\rm I\kern-.18em T}}
\def\ZZ{\relax{\sf Z\kern-.4em Z}}

\def\nablaslash{\relax{\rm /\kern-.28em \nabla}}

\def\a{\alpha}   \def\b{\beta}    \def\g{\gamma}  
\def\e{\epsilon} \def\G{\Gamma}     \def\l{\lambda}
\def\om{\omega}   \def\si{\sigma}

\def\cA{{\cal A}} 
   \def\cF{{\cal F}}
 \def\cH{{\cal H}} \def\cI{{\cal I}} 
\def\cK{{\cal K}} \def\cL{{\cal L}}  \def\cN{{\cal N}}
\def\cO{{\cal O}} \def\cP{{\cal P}} \def\cQ{{\cal Q}} \def\cR{{\cal R}}
\def\cS{{\cal S}}   \def\cV{{\cal V}}

\def\mathR{{\mathbb R}}     \def\mathZ{{\mathbb Z}}

 \def\tf{{\tilde f}} \def\tg{{\tilde g}}

\def\ttf{{\tilde{\tf}}}   \def\ttg{{\tilde \tg}}

    \def\bE{{\bar E}}    \def\bF{{\bar F}}

 \def\oE{{\overline E}}

 \def\oD3{{\overline \rmD 3}}

  \def\otau{{\overline \tau}}

\def\Ewhat{{\widehat E}}

\def\veck{{\vec{k}}}

\def\adot{{\dot{a}}}   
\def\Hdot{{\dot{H}}}        

   \def\phidot{{\dot{\phi}}}          \def\sidot{{\dot{\si}}}

\def\cRdot{{\dot{\cR}}}   \def\cSdot{{\dot{\cS}}}

\def\beq{\begin{equation}}
\def\eeq{\end{equation}}
\def\bea{\begin{eqnarray}}
\def\eea{\end{eqnarray}}
\def\llea#1{\label{#1}\eea}
\def\lleq#1{\label{#1}\eeq}
\let\nn=\nonumber

\def\notin{\ \hbox{{$\in$}\kern-.51em\hbox{/}}}
\def\notsubset{\ \hbox{{$\subset$}\kern-.63em\hbox{/}}}
\def\lra{\longrightarrow}

\def\del{\partial}

  \def\E1Fq{E_1/\IF_q}

\def\rmD{{\rm D}}

\def\rmCDM{{\rm CDM}}        \def\rmGL{{\rm GL}}

\def\rmSL{{\rm SL}}   

        \def\rmcorr{{\rm corr}}  
                    
   \def\rmiso{{\rm iso}}         
      \def\rmmod{{\rm mod}}
\def\rmnmod{{\rm nmod}}     \def\rmord{{\rm ord}}

      \def\rmtr{{\rm tr}}

  \def\rmRe{{\rm Re}}

 \def\rmD{{\rm D}}

\def\rmfl{{\rm fl}}

\def\rmmod{{\rm mod}}
     
             \def\rmord{{\rm ord}}

        \def\rmtr{{\rm tr}}

        \def\rmGeV{{\rm GeV}}   \def\rmGL{{\rm GL}}    
     
      \def\rmIm{{\rm Im}}     
   \def\rmMpc{{\rm Mpc}}

        \def\rmPl{{\rm Pl}}

\def\rmRe{{\rm Re}}         
   
\def\rmSL{{\rm SL}}

\def\notdiv{{\relax{~|\kern-.35em /~}}}

\def\boxit#1{
\vbox{\hrule height1pt\hbox{\vrule width1pt\kern0.3cm
\vbox{\kern0.3cm\hbox{$\displaystyle#1$}\kern0.3cm}\kern0.3cm\vrule
width1pt}\hrule height1pt}}

\begin{document}

\parindent=0pt

 \phantom{whatever  \hfill  \today~}

\vskip 1.2truein 

\centerline{\bf {\Large Modular Inflation at Higher Level $N$}}

\vskip .4truein

\centerline{\sc Monika Lynker\fnote{1}{mlynker@iusb.edu} and
                Rolf Schimmrigk\fnote{2}{rschimmr@iusb.edu, netahu@yahoo.com}}

\vskip .3truein

\centerline{Dept. of Physics}

\vskip .1truein

\centerline{Indiana University at South Bend}

\vskip .1truein

\centerline{Mishawaka Ave., South Bend, IN 46634}

\vskip 1truein
\baselineskip=18pt

\parskip=0pt
\centerline{Abstract}
\begin{quote}
We introduce the framework of modular inflation with level structure, generalizing the level one theory considered 
previously to higher levels.  We analyze the modular structure of CMB observables in this framework and show that
 the nontrivial geometry of the target space suffices to ensure the almost holomorphic modularity of the relevant 
 parameters.  We further introduce a concrete  class of models based on hauptmodul functions that provide 
 generators of the corresponding inflationary potentials at level $N>1$. The phenomenology of this class of 
 models provides targets for  ground-based CMB experiments in the immediate future. In the framework of our 
 models we also discuss the status of the quantum gravity conjectures that have been formulated in the context 
 of the swampland.
\end{quote}

\renewcommand\thepage{}
\newpage
\parindent=0pt

 \pagenumbering{arabic}

\baselineskip=17pt
\parskip=0.0pt

\tableofcontents

\vskip .4truein

\parskip=0.1truein
\baselineskip=18.4pt 

\section{Introduction}

Multifield inflation provides a useful  perspective for the experimental results obtained over the past years with 
CMB probes. While singlefield inflation and low-dimensional fits suffice to account for the data collected by the satellite
experiments  COBE, WMAP and {\sc Planck} \cite{b96etal, h12etal, planck13, planck15, planck18-6, planck18-10},
   such analyses do not give an accurate picture of what 
this data implies for the inflationary theory space. This is of particular importance because multifield theories predict 
 effects that are absent or suppressed in single field models and thereby
   provide guidance for the phenomenological analysis. More work is necessary to properly constrain with the currently
   available data   even the simplest extensions of inflationary models based on single fields.
   
 The inflationary landscape has remained a somewhat 
amorphous object in its most general framework because of the lack of an organizing principle.  Such a principle 
 has been introduced in refs. \cite{rs14, rs15} for a subclass of theories based on automorphic symmetries. 
 The automorphic structure
 endows the theory space with a foliation  and the inflationary potentials are restricted because their
  building blocks are automorphic forms \cite{b73,g06}. Each of the leafs of the foliation is specified by numerical 
  characteristics, and for each leaf only a finite number of fundamental building blocks exist.
 Furthermore,  each of these functions is determined by a finite amount of information, hence can be identified uniquely by
    a finite number of computations. A further motivation for automorphic inflation arises from the fact 
    that it provides a systematic framework in which the shift symmetry, often introduced as an ad hoc operation to provide 
    protection of the inflaton potential against corrections, is an element of a bona fide discrete group that arises as the remnant of a 
  weakly broken continuous group.
   This embedding of the shift symmetry allows to formulate the constraints on the model in an explicit and structured way.
 
 The emergence of discrete symmetries makes the framework of automorphic inflation more compatible with conjectures 
 concerning the nature of low energy theories that admit a UV completion. Inflationary models without symmetries will tend to be 
 less compatible. These conjectures, initiated in the papers \cite{v05, dl05} and  continued 
 in \cite{ov06,o18etal,agrawal18etal, gk18,oo18etal},  
 aim at distinguishing potentially  consistent theories from those that are not, 
 thereby relegating the latter to the swampland introduced by Vafa. While the main purpose in this paper  
 is not to add to the extensive discussions in the literature of these  conjectures,
 we will comment on the status of some of the conjectures in the context of automorphic and modular
  inflation as we describe  its structure in this paper.
 
The simplest realization of the above group theoretic formulation, leading to  inflationary models for which a symmetry
group  provides protection against the aforementioned corrections, is provided by modular inflation. 
 This represents the specialization to two fields of the more general theory of automorphic inflation, 
  which is defined in terms of an arbitrary number of inflaton components and in which the discrete symmetry group
   is a subgroup of a general higher rank reductive continuous symmetry group \cite{rs14, rs15}. 
 The idea to formulate automorphic inflation is analogous to the idea to  embed gauge theoretic duality transformations of 
 the coupling parameters into larger symmetry groups.  In the context of the full modular group 
there exists a distinguished generator of all modular invariant functions, the absolute invariant $j$. 
It is therefore natural to formulate a model of $j$-inflation based on this function. The phenomenological analysis  
 shows  that  this model is consistent with the data obtained from the CMB satellites  \cite{rs14,rs16},  and the reheating 
 epoch  after $j$-inflation was considered in \cite{rs17}. Subsequent work emphasized the hyperbolic geometry 
 underlying modular inflation in the context of 
     $\a$-attractors \cite{c1504etal, c1506etal,ls17, bl17, bl18},  and 
     hyperbolic inflation \cite{b17, mm17, bm19, f19etal}, 
while the discussion  of reheating after $j$-inflation was continued in ref. \cite{pv18}.

We will see below that the modular group is not the only discrete subgroup of the M\"obius group that contains
 the shift symmetry. In the present work 
 we introduce modular inflation at higher level congruence subgroups $\G_N$ of the full modular group, generalizing the 
 framework of modular inflation introduced in \cite{rs14, rs16} for the largest discrete subgroup of this type 
 in the continuous M\"obius group. 
 The picture thus is that the M\"obius group $G(\mathR)=\rmSL(2,\mathR)$ is weakly broken to some smaller subgroup  
 $\G_N \subset \rmSL(2,\mathZ)$ at a scale whose value is determined by the amplitude of the adiabatic power spectrum.
 The advantage of considering smaller symmetry groups is that they provide weaker constraints on the theory and
 therefore extend the reach of the modular  framework.
Such discrete subgroups at higher level also appear in the context of duality in 
    gauge theories \cite{k06, hm13, b15etal}, as well as in the  microscopic description of 
    black hole entropy \cite{s05, cs11}.
 In the present paper we consider such groups in the context of inflationary models. In principle one might consider 
 other classes of groups, such as Fuchsian groups for which a rich geometric theory exists that has been considered 
 in discussions concerned with attractor theory in \cite{ls17, bl17, bl18}. For such groups however
 the theory of modular forms and functions has not been developed sufficiently to lend itself to the framework of inflation
 with modular invariant potentials.
 
One of the issues that arise in inflationary models based on modular forms and functions
 is the modularity of the physical observables. Phenomenological 
variables such as the spectral index and the tensor-to-scalar ratio involve derivatives of the potential and hence derivatives
of the modular forms that are the building blocks of modular invariant inflaton potentials. Derivatives of modular forms
are not modular however, hence it is not a priori clear what the modularity properties are of the observables  
constructed from such derivatives. This problem has been addressed in the theory of 
 modular inflation associated to the full modular group in ref. \cite{rs16}. It was found there 
 that the curved nature of the field space serves to restore modularity,
albeit in a generalized sense, involving nonholomorphic modular forms.  In the present paper we generalize the 
 modularity analysis  of \cite{rs16}  to inflation at higher levels and general potentials. We furthermore make our general 
framework concrete by introducing a 
 class of inflationary modular invariant potentials given by hauptmodul functions that are based on
  modular forms relative to certain Hecke congruence subgroups.  The groups considered here are distinguished by the fact that
   the associated space of modular invariant inflation potentials have a single generator, much like in the case of the full 
   modular groups, where the $j$-function that  defines $j$-inflation is the generator for modular invariant functions relative to 
  $\rmSL(2,\mathZ)$ \cite{rs14, rs16}. 
We  further analyze the phenomenology of one of these models 
in more detail and establish that it leads to sufficient inflation with parameters that are compatible with 
the satellite constraints found by WMAP and {\sc Planck}.

This paper is organized as follows. In $\S2$ we briefly establish our multifield inflation notation and in $\S3$ we
 formulate the 
 multifield form of the CMB observables that we will focus on. In $\S$4 we introduce the general structure of higher 
 level modular inflation,
 describe in particular the general form of the CMB observables computed later in the paper and establish their modularity 
 in a generalized sense in terms of almost holomorphic modular forms. 
 In $\S$5  we  comment on the geometric part of the  quantum gravity conjectures in the context of modular inflation.
 In $\S6$ we briefly describe the transfer function evolution and 
 in $\S7$ we make our general framework concrete by introducing 
      the class of hauptmodul inflation models,  which we denote as $h_N$-inflation. 
 In $\S8,9$ we investigate  this class of models in some detail and 
 in $\S10$ we make some concluding remarks. 
  
  \vskip .2truein
  
 \section{Multifield inflation}
 
 In this section we briefly introduce our notation and recall a closed form for the 
  multifield inflation with curved target spaces, the framework  relevant for automorphic inflation in general and 
  modular inflation in particular. In this context the field space carries a Riemannian geometry given by a 
  nontrivial metric $G_{IJ}(\phi^K)$ with the associated 
  Levi-Civita connection determined by the Christoffel symbols $\G_{IJ}^K(\phi^L)$. The action is given in terms of 
  the Lagrangian
  \beq
   \cL ~=~ -\frac{1}{2}G_{IJ} g^{\mu \nu} \del_\mu\phi^I \del_\nu \phi^J
    ~-~ V(\phi^I).
  \eeq
  The perturbations $\delta \phi^I$ are encoded in terms of their covariant forms $\cQ^I$ introduced in \cite{gt11}
   via gauge invariant variables
  \beq
   Q^I ~=~ \cQ^I~+~ \frac{\phidot^I}{H} \psi,
  \eeq
  and projected to adiabatic and isocurvature directions. The adiabatic projection is chosen here to 
  be the comoving curvature perturbation $\cR$ of Lukash \cite{l80} and Bardeen \cite{b80}  
  (see also \cite{w08})
  which is given in the Newtonian gauge by
     \beq
    \cR ~=~  H\delta u ~-~ \psi ~=~  - \frac{H}{\sidot} \si_I Q^I.
  \eeq
 Here  $\psi$ is the spatial  perturbation of the Friedman-Lemaitre metric with the convention $\delta g_{ij} = 2a^2\psi \delta_{ij}$
 and $\sidot$ is a measure of the inflaton speed
  \beq
  \sidot ~:=~ \sqrt{G_{IJ} \phidot^I\phidot^J},
  \eeq
  while $ \si^I = \phidot^I/\sidot$
  defines the normalized velocity vector. $H=\adot/a$ is the Hubble-Slipher parameter, first determined with 
  Slipher's redshift data \cite{s15, s17, e24} and Hubble's distance data \cite{h29}.
 
The isocurvature perturbations have been encoded in a number of different ways in the literature. The approach taken in
 \cite{rs14,rs16} is to use the structures that source the dynamics of the adiabatic perturbation $\cR$ as it is  obtained 
  in the general 
case of multifield inflation with an arbitrary number of fields in the inflaton multiplet.  This dynamics has been derived  
in closed form for an arbitrary number of fields in \cite{rs16}
 in terms of the dimensionless contravariant fields
 \beq
  S^{IJ} ~=~ \frac{H}{\sidot}\left(\si^I Q^J - \si^J Q^I\right).
  \eeq
   and the covariant tensor 
   \beq
    W_{IJ} ~=~ \si_I V_{,J}- \si_J V_{,I}
   \eeq
 as 
\beq
 \cRdot ~= ~ \frac{H}{\Hdot} \frac{1}{a^2} \Delta \psi ~+~ \frac{1}{\sidot}  S^{IJ}W_{IJ}
 \eeq

  For large scales only the second term is relevant,  hence this source term motivates the identification
  of  the tensor $S^{IJ}$ as a measure for the isocurvature perturbations. The isocurvature projection 
  $S^{IJ}W_{IJ}/\sidot$, identified by the dynamics above as the large scale adiabatic source term, 
    is related to the contraction of the variables $Q^I$ with respect to the turn-rate  $\a^I$ often used 
    in the multifield literature. Different notions of a turn-rate have been considered and 
   the formal definitions that are in use are not all equivalent 
         \cite{g00etal, w02etal, t08etal, pt10, a10etal, pt11, mrx12, kms12, crs18, pr18,b18etal}.   
    Defining $\a^I = D_t\si^I$   and using the exact relation 
  \beq
   \a^I ~=~ D_t \si^I ~=~ - \frac{P^{IJ}V_{,J}}{\sidot}
   \eeq
   in terms of the projection operator $P^{IJ} = G^{IJ}-\si^I\si^J$ we can relate our source term above to the projection 
   $\a_IQ^I$ as 
   \beq
   \frac{1}{\sidot} S^{IJ}W_{IJ} ~=~ - 2\frac{H}{\sidot} \a_I Q^I.
  \eeq
  Since $\a^I$ is orthogonal to the trajectory unit vectors $\si^I$ it is natural to rescale $\a^I$ into a unit vector $\nu^I = \a^I/\a$
  with  $\a = \left|D_t\si^I\right|$.
   This leads to the dimensionless isocurvature projection
 \beq
  \cS ~:=~ \frac{H}{\sidot} \nu_I Q^I ~=~  - \frac{1}{2\a \sidot} S^{IJ}W_{IJ}.
   \lleq{isocurvature-proj-v2}
 In the case of modular inflation considered below the isocurvature tensor has only one independent component and the relation 
 (\ref{isocurvature-proj-v2}) can be viewed as a rescaling.
 
  In the slow-roll approximation the general multifield dynamics of the variables $(\cR, S^{IJ})$ for curved target spaces with an
   arbitrary number of   fields and associated adiabatic and isocurvature perturbations $S^{IJ}$ was obtained in 
   ref. \cite{rs16} in closed form as 
    \bea
     \cRdot &=& - 2H \eta_{IK}\si^K  \si_J S^{IJ}\nn \\
   D_t  S^{IJ} &=&  2H\left(\eta_{KL}\si^K\si^L-\e\right) S^{IJ}  +
    H\left( \eta_{KL} G^{K[I} S^{J]L} 
      - \frac{\e}{3}M_\rmPl^2 \si^{[I}R^{J]}_{KLM}\si^K S^{LM}\right),
    \llea{mfi-sr-dynamics}
   where $\e =- \frac{\Hdot}{H^2}$ is expressed in the slow-roll approximation in terms of slow-roll parameters 
   \beq
    \e_I ~=~ M_\rmPl \frac{V_{,I}}{V}
    \eeq
    as $\e=G^{IJ}\e_I\e_J/2$ 
    and $\eta_{IJ}$ are further slow-roll parameters, defined here as 
   \beq
    \eta_{IJ} ~=~ M_\rmPl^2 \frac{V_{;IJ}}{V}.
   \eeq
   For the Levi-Civita connection this defines a symmetric tensor.
    The components $R^I_{JKL}$ of the curvature tensor are defined in terms of the Levi-Civita 
   connection of the field space metric $G_{IJ}$. Here the antisymmetrization is defined as 
   $\si^{[I}W^{J]} = \si^I W^J - \si^JW^I$.   This generalizes the flat target twofield dynamics of ref. \cite{w02etal} to an arbitrary number of
  fields and curved field spaces. 

\vskip .2truein

\section{CMB observables in MFI} 

The most important CMB observable for the experiments in the recent past has been the spectral index.
 In general, multifield inflation with curved target spaces this can be written in various ways. 
 The dimensionless power spectrum $\cP_{\cR\cR}$, defined as 
\beq
 \cP_{\cR\cR} ~=~ \frac{2\pi^2}{k^3} P_{\cR\cR} 
 \eeq
 with
 \beq
 \langle \cR(\veck) \cR(\veck')\rangle ~=~ (2\pi)^3 \delta^{(3)}(\veck + \veck') P_{\cR\cR}(k)
\eeq
can be written in terms of the Hubble-Slipher parameter 
and   can be expressed in the slow roll approximation in terms of the potential as 
  \beq
   \cP_{\cR\cR} ~=~ \left(\frac{H}{2\pi}\right)^2 \left(\frac{H}{\sidot}\right)^2 
    ~=~ \frac{1}{12\pi^2 M_\rmPl^4} \frac{V}{G^{IJ}\e_I\e_J}.
   \eeq
 With the slow-roll parameters $\e_I$ and $\eta_{IJ}$ the spectral index defined for perturbations $\cO, \cO'$ as
 \beq
 n_{\cO\cO'} ~=~ 1 ~+~ \frac{d\ln \cP_{O\cO'}}{d\ln k} 
 \eeq
 takes for $\cO=\cR=\cO'$ the form
 \beq
  n_{\cR\cR} ~=~ 1 - 3G^{IJ} \e_I\e_J ~+~ 2 \frac{\eta_{IJ}\e^I\e^J}{G^{KL}\e_K\e_L}.
 \lleq{mfi-nRR}

Associated to  the isocurvature perturbations $S^{IJ}$ are the power spectra $\cP_{\cR \cS^{IJ}}$ and $\cP_{\cS^{IJ} \cS^{KL}}$
and their spectral indices $n_{\cR \cS^{IJ}}$ and $n_{S^{IJ}S^{KL}}$, 
which can be explicitly computed as well. The power spectrum $\cP_{\cS\cS}$ of the isocurvature 
projection (\ref{isocurvature-proj-v2}) coincides with that of $\cP_{\cR\cR}$ at
horizon crossing
   \beq
   n_{\cS\cS} ~=~ n_{\cR\cR}
  \eeq
and there is no cross correlation $\cP_{\cR\cS}=0$. Hence at horizon crossing the correlation coefficient $\g_\rmcorr$, defined by 
 \beq
  \g_\rmcorr ~=~ \frac{\cP_{\cR\cS}}{\sqrt{\cP_{\cR\cR}\cP_{\cS\cS}}},
  \eeq
  vanishes, and the isocurvature ratio $\b_\rmiso$, defined as
  \beq
  \b_\rmiso ~= \frac{\cP_{\cS\cS}}{\cP_{\cR\cR}+\cP_{\cS\cS}}
 \eeq
 evaluates to $\b_\rmiso=1/2$.

 The primordial gravitational contribution of the power spectrum has not been observed directly. It is conventionally 
  parametrized in terms of the  tensor-to-scalar ratio $r$ defined as $r = \cP_T/\cP_{\cR\cR}$, where the 
  tensor power spectrum is given by \cite{s79} 
  \beq
   \cP_T ~=~ \frac{2}{\pi^2} \left(\frac{H}{M_\rmPl}\right)^2.
  \eeq
  Both WMAP and {\sc Planck} have strengthened the bounds on $r$ considerably over the past decade, with important 
  ramifications for model selection. In multifield inflation the total scalar 
  power spectrum includes  contributions of the isocurvature perturbations, but we retain the above definition, which can be written 
  in terms of the slow-roll parameters as
  \beq
  r ~=~ \frac{8}{M_\rmPl^2} \left( \frac{\sidot}{H}\right)^2 ~=~ 8 G^{IJ}\e_I \e_J.
  \lleq{r-ten}
 The tensor spectral index, commonly defined without the  shift of $(-1)$ conventionally adopted for the scalar power spectra as
 \beq
  n_T ~=~ \frac{d\ln \cP_T}{d\ln k}, 
  \eeq
  is not an independent parameter but is given in the slow-roll approximation by 
  \beq
   n_T ~=~ - G^{IJ}\e_I\e_J ~=~ - \frac{r}{8},
   \eeq
   leading to a red-tilted spectrum. These relations are modified by the transfer functions considered further below.
 
 The number of e-folds is a further key parameter that has to be considered. There are a number of different ways to encode this, with
 the most common being the number of e-folds  $N_k$ between the horizon crossing of $k$ and the end of inflation
  \beq
   N_k ~=~ \int_{t_k}^{t_e} H dt.
   \eeq
 In multifield inflation $N_k$ has to be evaluated numerically in general, after solving the dynamics. In the context of 
 slow-roll inflation the first Friedman-Lemaitre equation transforms this into an integral that only involves the 
 inflaton potential $V(\phi^I(t))$.
 
 In twofield inflation there are then, in principle, at least 12 parameters of relevance for the experimental collaborations to fit, given 
 by the amplitudes and indices of the power spectra $(A_{\cO\cO'}, n_{\cO, \cO'})$ for $\cO, \cO'=\cR, \cS$, the tensor observables
 $(A_T, n_T)$, and  the  canonical cosmological parameters given by the density parameters $\om_\rmCDM, \om_B, \om_\Lambda$ 
 and the optical depth $\tau$. In the slow-roll approximation these parameters reduce at horizon crossing according to the above 
 relations.  
 
 Recent work in inflation has been concerned with  conjectures initiated some time ago in refs. \cite{v05, dl05, ov06}, 
 and developed  further in \cite{o18etal, agrawal18etal,oo18etal},  to the effect that theories that can be extended
  in a UV consistent way should satisfy a number of criteria that can be made semi-quantitative. 
 The most recent among these takes the form of a gradient conjecture, which posits that the relative dimensionless 
 gradient of the potential should be bounded from below by a positive constant \cite{o18etal}. In the original conjecture, 
 made in the context of a single scalar field with a flat metric, this constant is assumed to 
 be of order one, in which case one obtains a conflict with the slow-roll condition of single field inflation. 
 The above discussion shows how more generally the geometry of the multifield target space interferes when relating 
 the gradient of the 
 potential to the observables of the theory, such as the tensor-to-scalar ratio $r$. We will comment further below on the 
 quantum gravity conjectures  in the context of modular inflation.
 
\vskip .2truein

\section{Modular inflation at higher level $N>1$}
 
 Modular inflation is the simplest framework in which the shift symmetry is an element of a discrete group. The formulation
 introduced in \cite{rs14,rs16} is concerned with the modular group $\rmSL(2,\mathZ)$ 
  because it provides the strongest possible constraints in the standard theory of modular forms.
 A special feature of this  framework is that the space of all modular invariant function is particularly simple because 
 it is  generated by a single function, the absolute invariant, denoted by $j$, 
 in the sense that all others can be expressed in terms of $j$ by a quotient of two 
 polynomials in $j$. The $j$-function  thus is a distinguished element in this space and it is natural to use it to construct a modular invariant 
 potential.  It was shown in refs. \cite{rs14,rs16,rs17} that the resulting inflationary model is compatible 
 with the CMB observations of WMAP and {\sc Planck}. Reheating aspects of this model were furthermore discussed in \cite{rs17, pv18}.
 In the present section we generalize the framework considered in \cite{rs14, rs16} to  subgroups at higher levels $N>1$.
 
 \subsection{Breaking the M\"obius group to congruence subgroups}
 
 The full modular group is not the only discrete subgroup of the M\"obius group $\rmSL(2,\mathR)$ that contains the 
 shift symmetry and it is natural to explore whether the modular structure of $j$-inflation can be generalized to subgroups $\G$ of 
  $\rmSL(2,\mathZ)$. In this more general context the continuous M\"obius group is 
  weakly broken to a discrete subgroup $\G \subset \rmSL(2,\mathZ)$, thereby realizing a special case of the symmetry
 breaking $G(\mathR) \lra  G(\mathZ)$ outlined in \cite{rs14,rs15} for general reductive groups $G$. Smaller groups will provide 
  weaker constraints, thereby enhancing the possibilities for modular 
 invariant potentials. The goal in this paper is to focus on groups that have the same generating property as the modular group in the 
 sense that their constraints are strong enough to identify a single generator of the associated space of modular invariant functions,
 leading to natural candidates for the inflationary potentials.  For $\rmSL(2,\mathZ)$  the existence of the special element $j(\tau)$ is 
 explained by the fact that the compactified form of the quotient $\cH/\rmSL(2,\mathZ)$ of the complex upper halfplane $\cH$ is 
 a complex curve of genus zero, i.e. a sphere. While in general discrete subgroups $\G$ of $\rmSL(2,\mathR)$ 
 lead to higher genus curves, in which case the modular invariant functions have more than one generator, there exist subgroups 
 which contain the shift symmetry and which 
 lead to the genus zero property.  
 
 Because we are interested in building blocks of  inflationary potentials that are modular forms
 we focus here on the class of congruence subgroups, defined as those groups that contain a principal congruence
  subgroup $\G[N]$  given by those elements $\g \in \rmSL(2, \mathZ)$ that are congruent to the unit matrix ${\bf 1}$ mod $N$
 \beq
  \g ~\equiv ~ {\bf 1} (\rmmod~N),
 \eeq
  where the level $N$ is a positive integer.   Among  these congruence subgroups the most important type is given by the 
  Hecke subgroups $\G_0(N)$ of level $N$, defined by matrices
  \beq
   \g ~=~ \left(\begin{matrix} a &b\cr c &d\cr\end{matrix} \right)
  \eeq
  such that $c$ is a multiple of $N$. These groups have been the main focus in the theory of modular forms for more than a century,
  leading to the most detailed structure for these objects. 
    In the following we will focus on this framework.  In the class of  Hecke congruence subgroups there are a finite number
    of levels for which the complex curve associated to the group is of genus zero, leading to generators that
  we will denote by $h_N$. These generators are the natural generalization of the 
  $j$-function and form the basis for our formulation of $h_N$-inflation introduced below. For Hecke groups
  $\G_0(N)$ the specific levels that have genus zero property are given by 
  $N=1,...,10$, 12, 13, 16, 18, 25 \cite{s74}. 
 
 In a general modular  inflation model  we take the potential to be given by 
 \beq
  V ~=~ \Lambda^4 |F(f_i)|^{2m}
 \lleq{F-potentials}
 where $F$ is a dimensionless function of the modular forms $f_i$ associated to congruence groups 
 $\G_i\subset \rmSL(2,\mathZ)$ of the full modular group and  $m$ is an integer.
  Thus for $\g_i = {\tiny \left(\begin{matrix}  a &b\cr c &d\cr \end{matrix}\right)} \in \G_i$ the $f_i$ transform as 
  \beq
   f_i(\g_i \tau) ~=~ \chi_i(\g) (c\tau+d)^{w_i} f(\tau),
  \eeq
  where $w_i$ is the weight of $f_i$ and $\chi_i$ is a possibly trivial character.
   The phenomenological analysis of modular inflation models involves derivatives of the potential and 
   hence derivatives of the modular forms that define the model. Such derivatives in general involve nonmodular
 functions, which raises the question of 
 the modularity status of inflationary observables. This issue was addressed first in \cite{rs16} for modular 
 inflation determined by modular forms relative to the full modular group. The fundamental picture in that case
 can be summarized by noting that the set of generators of the space of all modular forms has to be 
 enlarged in the space of modular forms and their derivatives by including the nonmodular Eisenstein series 
 $E_2$. This series transforms under a modular transformation $\g \in \rmSL(2,\mathZ)$ of the full group as
 \beq
  E_2(\g \tau)~=~ (c\tau+d)^2 E_2(\tau) - \frac{6i}{\pi} c(c\tau+d),
 \eeq
 which makes it reminiscent to the inhomogeneous behavior of the Christoffel symbols. It was shown in \cite{rs16} however that 
 the CMB observables $n_{\cR\cR}$ and $r$ are modular invariant because the terms induced by the nontrivial geometry 
 of the field space combine with the $E_2$-terms into modular invariant terms. These involve the modified Eisenstein series
\beq
 \Ewhat_2(\tau) ~:=~ E_2(\tau) ~-~ \frac{3}{\pi \rmIm~\tau},
\lleq{Ewhat2}
which does define a modular form, albeit a nonholomorphic one.
 In the more general case of modular forms with respect to congruence subgroups $\G_N \subset 
 \rmSL(2,\mathZ)$ the symmetry constraints are weaker and the resulting structure is different. Since we will need 
 to consider both level one and higher level modular forms for our models we briefly recall the pertinent points for the 
 level one case from \cite{rs16} before discussing the higher level generalization. 

\subsection{Derivatives of the inflaton potential}

 For the spectral index of the scalar power spectrum and the  gravitational contribution of the total power spectrum, 
 as measured by $r$,  it is sufficient to know the first and second derivatives  of the potential.
   Writing the modular function $F$ of the inflaton potential as a quotient
  \beq
   F ~=~ \frac{f}{g}
  \eeq
  of two modular forms $f,g$ with characteristics $(w_f, N^f, k_f)$ and $(w_g, N^g, k_g)$ respectively. Here 
  the integer $w_f$ denotes the weight of the form, $N^f$ is its level, and $k_f$ is the lowest order in its 
  Laurent expansion
   \beq
    f(q) ~=~ \sum_{n\geq k_f} a_n q^n,
   \eeq
   and similarly for $g$. If $w_f=w_g$ the function $F$ is a weight zero function. 
  
For modular forms of weight $w$ and level $N$ the derivative furthermore involves higher level Eisenstein series, 
which we denote here by $E_w^N$, defined as  
  \beq
  E_w^N(\tau) := E_w(N\tau),
  \eeq
  where $E_w(\tau)$ is the standard Eisenstein series of weight $w$ given by 
  \beq
  E_w(\tau) ~=~ 1 - \frac{2w}{B_w} \sum_{n\geq 1} \si_{w-1}(n) q^n
 \eeq
 where $q=e^{2\pi i \tau}$, the denominators $B_w$ are the Bernoulli numbers, and 
 \beq
  \si_w(n)~=~ \sum_{d|n} d^w
  \eeq
  is the divisor function.  For $w>2$ the Eisenstein series are modular forms of weight $w$ with respect to the full modular 
  group, and the series defined by $E_w^N$ are modular forms with respect to a congruence subgroup 
  of level $N$.
  
  To determine the derivatives of modular forms relative to the Hecke congruence subgroups is somewhat involved and 
  has only recently been  completed for the groups of genus zero that are of interest in the present paper \cite{a02, a05, gr11}.
   In the interest of simplicity we focus in the following on groups of prime level, even though the analysis can be 
   generalized to the remaining groups at the cost of more complicated formulae.  
   In the case of prime level the derivatives can be 
   written as
  \beq
   f' ~=~ 2\pi i \left(\tf + \frac{k_f - \frac{w_f}{12}}{p_f-1} f (p_fE_2^{p_f} - E_2) 
      + \frac{w_f}{12} f E_2\right), 
 \lleq{fprime}
 where  $\tf$ is a modular form of weight $(w+2)$. The second term also defines a  modular form of the same weight, 
 as can be seen from the  transformation behavior of  $E_2^N$, which is given by
 \beq
  E_2(N\g \tau) ~=~ (c\tau+d)^2 E_2(N\tau) ~-~ \frac{6ic}{N\pi} (c\tau+d), ~~~~
  \g ~= ~ \left(\begin{matrix} a &b\cr c&d\cr \end{matrix}\right) \in \G_0(N).
 \eeq
 The non-modular second term here is cancelled by the corresponding term of $E_2$, hence $NE_2^N - E_2$ defines 
 a modular form of weight two for $\G_0(N)$. Since this combination of $E_2^N$ and $E_2$ appears throughout our
 analysis, it is useful to introduce the abbreviation 
  \beq
   E_{2,N} ~:=~ NE_2^N ~-~ E_2.
  \lleq{E2levelN}
As noted above, the Eisenstein series $E_2$ is not modular, hence  the final term of $f'$ is  not modular. We will 
return to this further below in the context of the CMB observables. 
 An analogous expression is obtained for $g'$.  For modular forms at level $N=1$ this result specializes to 
  \beq
 f' ~=~ 2\pi i \left(\tf ~+~ \frac{w}{12} f E_2\right),
 \eeq
 for which the explicit construction of $\tf$ at level one has been described already in ref. \cite{rs16}. 
 
 The modular form of weight two $\tf$ is determined in the present case in terms of the quotient 
 \beq
 h_p(\tau) ~=~ \left(\frac{\eta(\tau)}{\eta(p\tau)}\right)^{24/(p-1)} 
 \eeq
  as 
  \beq
   H_p(\l, \tau) ~:=~ \frac{E_{2,p}(\tau)}{(p-1)} \left( \frac{h_p(\tau)}{h_p(\tau) - h_p(\l)} ~-~ 1\right).
  \eeq
 Here $\eta(\tau)$ is the  Dedekind eta-function, which 
  can be written as an infinite product in terms of the 
 variable $q=e^{2\pi i\tau}$ as 
 \beq
 \eta(q) ~=~ q^{1/24} \prod_{n\geq 1} (1-q^n).
 \lleq{dedekind-eta}
  Given these ingredients the form $\tf$ associated to $f$ is given by the following sum over the 
  fundamental domain $\cF_p$ of $\G_0(p)$
  \beq
   \frac{\tf}{f} ~=~ - \sum_{\l \in \cF_p} \frac{1}{o_\l^p}\rmord_\l(f)~ H_p(\l, \tau),
  \lleq{ftilde}
  where $\rmord_\l(f)$ is the vanishing order of $f$ at $\l$ and $o_\l^p$ is the order of isotropy groups at $\l$.
  More details about the structure of $H_p(\l, z)$ can be found in ref. \cite{a02}.

With these ingredients we find that the first derivative is modular, and the quotient $F'/F$ that appears in the 
CMB observables   is of the form
  \beq
    \frac{F'}{F}
    ~=~ 2\pi i \left(\frac{\tf}{f} - \frac{\tg}{g} \right) 
      ~+~ 2\pi i \left(\frac{k_f-\frac{w}{12}}{p_f-1} E_{2,p_f}
                       ~-~ \frac{k_g-\frac{w}{12}}{p_g-1} E_{2,p_g}
      \right).
\lleq{Fprime-overF}
This quotient  determines the slow roll parameters  $\e_I$ explicitly in terms of the modular forms $f, \tf$, the Eisenstein 
series $E_2^N$, and $E_2$.  The fact that this expression shows the quotient $F'/F$ to be
a modular form of weight two will have implications for the general structure
of inflationary observables discussed below.
   
 For the spectral indices of the inflationary power spectra the second order derivatives are necessary. 
For higher level modular functions $F$ these second derivatives can be obtained by iteration in terms of modular 
 forms $\ttf, \ttg$ of weight $(w+4)$ that arise from the derivatives of $\tf$ and $\tg$.  
Introducing the abbreviations
\beq
    C_{p_f} ~:=~  (2\pi i) \frac{k_\tf-\frac{w_\tf}{12}}{p_\tf-1}  ~\frac{\tf}{f} E_{2,p_f}
                  +  \left(\frac{F'}{F} - 2\pi i \frac{\tf}{f} \right)   \frac{k_f-\frac{w}{12}}{p-1} E_{2,p_f} 
                  +\frac{( 2\pi i)}{12} \frac{k_f-\frac{w}{12}}{p-1} \left(  \left(E_{2,p_f} \right)^2 -\left(E_{4,p_f}\right)\right)
 \eeq
and similarly for $C_{p_g}$ we find for the second 
derivative the weight four expression
\beq
 \frac{F''}{F}
  = 2\pi i \frac{F'}{F} \left( \frac{\tf}{f}-\frac{\tg}{g}\right) +(2\pi i)^2  \left( \frac{\ttf}{f} -  \frac{ \ttg}{g} \right)
        -  (2\pi i)^2 \left( \frac{\tf^2}{f^2} -\frac{\tg^2 }{g^2} \right) 
         ~+~ 2\pi i\left(C_{p_f} - C_{p_g}\right) ~+~  \frac{\pi i}{3} \frac{F'}{F} E_2.
 \lleq{Fdoubleprime-overF}
This decomposes into a modular part $(F''/F)_\rmmod$ given by the first four terms and the nonmodular 
part given by the last term 
\beq
 \left( \frac{F''}{F}\right)_\rmnmod ~=~ \frac{\pi i}{3} \frac{F'}{F} E_2.
 \lleq{nonmodular-part}
 This will be enter below in the structure of the scalar spectral indices.

\vskip .1truein

\subsection{Geometry of higher level field spaces}

As indicated earlier in the general multifield discussion, the observables in inflation depend on the geometry of 
the target field space, which in modular inflation is given by the Poincar\'e metric
 \beq
  ds^2 ~=~ \frac{(d\tau^1)^2 + (d\tau^2)^2}{(\tau^2)^2} ~=~ \frac{d\tau d\otau}{(\rmIm~\tau)^2}
  \eeq
  with the K\"ahler potential
    \beq
   K ~=~ -4\ln (-i(\tau-\otau))
  \eeq
  and the nonvanishing Christoffel symbols 
   \beq
    \G_{11}^2 ~= ~ - \G_{22}^2 ~=~ - \G_{12}^1 ~=~ \frac{1}{\mu (\rmIm ~ \tau)}.
   \eeq
  Also needed is the curvature tensor, which in two dimensions has only 
   one independent component, given in modular inflation by 
   \beq
    R_{1212} ~=~ - \frac{1}{\mu^2 (\rmIm~\tau)^4}.
    \eeq
 The curvature scalar $R = G^{IJ}R_{IJ}$ of this curvature tensor is constant
 \beq
  R ~=~ - \frac{2}{\mu},
 \eeq 
 leading to the Gaussian curvature $\cK = - 1/\mu^2$. This shows that the energy scale $\mu$ in our formulation
 of modular inflation is a direct measure of the curvature of the field space. This scale, and hence the nontrivial 
 target space geometry enters in the CMB observables of  all modular inflation models.
  
While the metric of the field spaces is independent of the level $N$, the spaces themselves 
do depend on the level because the fundamental domain that defines the irreducible region of field space that 
generates the complete upper halfplane 
by motions via the discrete groups is given by the quotient
 \beq
  \cF_N ~:=~ \cH/\G_N.
 \eeq
 The specific form of these regions can be obtained more explicitly 
  from the fundamental region $\cF=\cF_1$ of the full modular group $\rmSL(2,\mathZ)$ 
 via a finite union of images of $\cF$ under maps constructed from the basic generators of the group.

\vskip .1truein

\subsection{Modularity properties of CMB variables}

The quotient $F'/F$ completely determines the tensor-to-scalar ratio and is part of the spectral index
 because the parameters $\e_I$ are given for potentials of type (\ref{F-potentials}) by
 \beq
  \e_I ~=~ i^{I-1} \frac{M_\rmPl}{\mu} \left(\frac{F'}{F} - (-1)^I \frac{\bF'}{\bF}\right),
 \eeq
 which leads with eq. (\ref{r-ten})  to 
 \beq
  r ~=~ 32\left(\frac{M_\rmPl}{\mu}\right)^2 (\rmIm~\tau)^2 ~\left|\frac{F'}{F}\right|^2,
 \lleq{modular-r}
hence the tensor-to-scalar ratio is given in terms of the fundamental building blocks by
\beq
 r ~=~ 128\pi^2 \frac{M_\rmPl^2}{\mu^2} (\rmIm~\tau)^2 \left| \left(\frac{\tf}{f} - \frac{\tg}{g}\right)  
      + \frac{k_f - \frac{w}{12}}{p-1} E_{2,p_f} ~-~ \frac{k_g - \frac{w}{12}}{p-1} E_{2,p_g}\right|^2.
 \eeq
 The analysis of the subsection \S4.2, in particular eq. (\ref{ftilde}), makes
 the modular invariance of $r$ in terms of standard holomorphic modular forms explicit.

The spectral index involves both the first derivative parameters $\e_I$  and the second derivative 
 slow-roll parameters $\eta_{IJ}$, which are defined in modular inflation via the covariant derivative 
because of the curved structure of the target field space. It is useful to decompose the parameters 
$\eta_{IJ}$ into a flat and a curved part as
\beq
\eta_{IJ} = \eta_{IJ}^\rmfl+ \eta_{IJ}^\G,
\eeq
 where the flat part is given by 
 \beq
  \eta_{IJ}^\rmfl ~=~ -  \frac{M_\rmPl^2}{\mu^2} i^{I+J} \left(\frac{F''}{F}
         ~-~ \left((-1)^I + (-1)^J \right) \left|\frac{F'}{F}\right| + (-1)^{I+J} \frac{\bF''}{\bF}\right)
  \eeq
 and  the  metric induced term $\eta_{IJ}^\G$ is given by
  \beq
   \eta_{IJ}^\G ~=~ - \frac{M_\rmPl^2}{\mu^2} \frac{i^{I+J-1}}{\rmIm~\tau} \left(\frac{F'}{F} 
         ~-~ (-1)^{I+J} \frac{\bF'}{\bF}\right).
 \eeq
   With these parameters the multifield spectral index (\ref{mfi-nRR}) can be computed to be 
   of the form
 \beq
  n_{\cR\cR} ~=~  1- 4 \frac{M_\rmPl^2}{\mu^2} (\rmIm~\tau)^2 \left[ 2 \left|\frac{F'}{F}\right|^2  -  
      \rmRe\left( \frac{F''}{F'} \frac{\bF'}{\bF}\right)\right]
          ~-~ 4 \frac{M_\rmPl^2}{\mu^2} (\rmIm~\tau) \rmIm\left(\frac{F'}{F}\right),
 \lleq{F-spectral-index-nRR}
where $F'/F$ is as in eq. (\ref{Fprime-overF}) and $F''/F'$ is obtained from 
eq. (\ref{Fdoubleprime-overF}) and eq. (\ref{Fprime-overF}).

The results of the previous subsection 
in combination with the expression (\ref{F-spectral-index-nRR}) make it possible to analyze the modularity of 
the spectral index in a concise way even though the detailed structure in terms of the fundamental building 
blocks at levels $N>1$ is somewhat involved. The above formula shows that the nonmodular term in $F''/F$ 
determined  in eq. (\ref{nonmodular-part})  contributes a nonmodular term to the
 spectral index that contains the factor
   \beq
   \left(\frac{F''}{F'}\right)_\rmnmod  ~=~  \frac{\pi i}{3} E_2.
  \eeq
  This nonmodular part of $F''/F'$ can be combined with the metric-induced final term of (\ref{F-spectral-index-nRR})
  to lead to the  modified Eisenstein series $\Ewhat_2$ defined in eq. (\ref{Ewhat2}).
  It follows from the transformation behavior of $\rmIm(\tau)$ given by
 \beq
  \frac{1}{ \rmIm ~\g \tau} ~=~ \frac{(c\tau+d)^2}{\rmIm ~\tau} + 2ic(c\tau+d) ~=~ \frac{|c\tau+d|^2}{\rmIm~\tau}
 \eeq
 that this modified Eisenstein series is a modular form of weight two. Thus, while $\Ewhat_2$ is not holomorphic,
 it is modular. As a result the final two terms in (\ref{F-spectral-index-nRR}) lead 
 to a modular invariant spectral index at higher level $N$ given by
 \beq
  n_{\cR\cR} ~=~  1 - 4 \frac{M_\rmPl^2}{\mu^2} (\rmIm~\tau)^2 \left[ 2 \left|\frac{F'}{F}\right|^2  -  
      \rmRe\left( \frac{F''}{F'} \frac{\bF'}{\bF}\right)_\rmmod 
      + \frac{\pi}{3}  \rmIm\left( \Ewhat_2~  \frac{\bF'}{\bF}\right) \right].
  \lleq{F-spectral-index-modinv}
  
  We see from this  that, as in the case of modular inflation at level one,  the spectral index for higher level 
 modular inflation is modular, although it is modular in a weaker sense because it involves an 
 almost holomorphic modular forms.
 (More background and references concerning quasimodular and almost holomorphic  modular forms can be found 
 in ref. \cite{rs16}.)
 This generalized modularity is again obtained via the contribution of the curved target space that arises through the 
 Christoffel term induced by the 
 covariant derivative that enters the definition of the parameters $\eta_{IJ}$. This term combines with the quasimodular 
 term in the spectral index into the almost modular form $\Ewhat_2$.   What changes at higher level is the specific form 
 of the spectral index in terms of the fundamental variables given by the modular forms $f,g$ that define the inflaton 
 potential, and the modular forms $\tf, \tilde{\tf}, \tg, \ttg, E_{2,p}, E_4$ that appear through the derivatives of the potential.
 
\vskip .2truein

\section{Geometric quantum gravity conjectures and modular inflation at higher level}

We have noted above that in the past few years attempts have been made to formulate in concrete ways  
the issue whether inflation is constrained  by quantum gravitational considerations in the sense that as an effective field theory 
 it should be in principle be possible to embed it into a theory that admits a UV completion. Several conjectures have been 
suggested that formulate semi-quantitative expectations about what might distinguish theories that are UV completable 
from those that are not, the latter forming the swampland. Most of these constraints are concerned with moduli theory, 
hence are not directly relevant for inflation.  It is however of interest to 
consider the geometry of the field theory space in a more general context, and in the present discussion we analyze
 how  the geometric structure of modular inflation for arbitrary level $N$ relates 
to the geometry conjectured to be imposed by quantum gravity on the field space of scalar field theories. 
Perhaps the earliest conjecture in this direction is the volume conjecture for the field space, introduced in the discussion  
of the swampland by Vafa \cite{v05}, and in a different context in \cite{dl05}. In the general framework this constraint 
states that  the $n$-dimensional volume obtained from the target space metric $G_{IJ}$  should be finite.  A second 
conjecture formulated in the swampland discussions is the diameter conjecture. In the present section we consider these 
geometric conjectures for theories of modular inflation at higher level $N$.

For modular inflation at level one the area of the fundamental domain is given by $\pi/3$, but for higher $N$ the
 fundamental domain is larger because the group theoretic constraints are weaker.  For modular inflation at 
 arbitrary level $N$ for any discrete subgroup $\G_N$ in $\rmSL(2,\mathZ)$ 
 the fundamental domain takes the form $\cF_N = \cH/\G_N$ and the volume conjecture
  takes for the complexified inflaton doublet $\tau$ the form
 \beq
  \int_{\cF_N} ~\frac{d\tau d\otau}{(\rmIm~\tau)^2} ~<~\infty.
 \eeq
 One way to think about $\cF_N$ is in terms of elements $\g_i$ in $\rmSL(2,\mathZ)$ such that the combination of the 
  sets  $\g_i \G_N$ recovers the modular group.
   With these elements the fundamental domain of $\G_N$ can be constructed as  
   \beq
    \cF_N ~= ~\bigcup_{i=1}^{\mu_N} \g_i^{-1} \cF,
   \eeq
   where the index $\mu_N =[\rmSL:\G_N]$ denotes the number of sets $\g_i \G_N$ needed to obtain $\rmSL(2,\mathZ)$. 
  For the Hecke groups $\G_0(N)$  this  index can be obtained in terms of the prime divisors $p|N$ of the 
  level $N$,  leading to the volume of $\cF_N$ given by 
 \beq
 \cV(\cF_N) ~=~ \frac{\pi}{3}  N \prod_{p|N}\left(1+\frac{1}{p}\right).
 \eeq
  This shows that the volume conjecture holds for all modular inflation models at arbitrary level  $N$. It also shows that the 
  discrete symmetry group motivated by the shift symmetry is necessary to obtain this constraint: without  it the volume of 
  the upper halfplane with the volume element given by the Poincar\'e metric would be infinite. In multifield inflation the 
  target spaces that have been considered in the literature are infinite already in the special case of two fields. 
  In such cases the most natural way to reduce the 
  target space is by considering quotients of the field space by a discrete symmetry. If the resulting domain is unbounded 
  in some directions then the quotient construction is still not enough if the target space is chosen to be flat, which is 
  the case in most of the multifield literature. Compliance with the volume conjecture is facilitated by both the existence 
  of a discrete group and a nontrivial metric, and  inflationary theories without these ingredients are under pressure from 
  the volume conjecture.
 
 A second conjecture that has been formulated in the context of UV completable theories in ref. \cite{ov06} is the 
 diameter conjecture. 
 In the context of moduli this conjecture claims that consistent theories are distinguished by  the existence of paths 
 in the target space 
 that have infinite length in the metric $G_{IJ}$ (see also \cite{h17etal}).
 The above construction of the fundamental domain $\cF_N$ shows that the domain $\cF$ of the full modular group is part 
 of $\cF_N$, as expected from the fact that the Hecke subgroups $\G_0(N)$ are strict subgroups. Since $\cF$ contains paths of 
 infinite lengths, in particular along the imaginary axis, so do the domains $\cF_N$. While the Poincar\'e metric decreases in the direction 
 of increasing imaginary  values, the distance diverges logarithmically. It follows that the diameter conjecture is also satisfied in 
 modular inflation at arbitrary levels $N$.  For more general theories with potentials it is natural to consider the inflaton 
 trajectories on the potential surface, in which case the target space metric $G_{IJ}$ is modified \cite{rs18}. This modified 
 metric still leads to paths of infinite length, hence the appropriate version of the diameter conjecture holds in the present 
 framework. 

 A final geometric conjecture about the target space of low energy UV consistent  theories is less of geometric 
 but of topological nature. It has been posited in the work of Ooguri and Vafa \cite{ov06} that the field space should be simply connected.
 The above construction of the fundamental domain in terms of the images of $\cF$ under the elements that 
 generate the left cosets can be used to show that the regions $\cF_N$ are in fact 
  simply connected.
 
 We therefore see from this discussion that all three quantum gravity conjectures of geometric type whose violation would place a
  theory in  the swampland are satisfied in all modular inflation theories at arbitrary level $N$. The essential features that bring 
 modular inflation in compliance with these conjectures is the existence of an infinite discrete symmetry group and a curved
 field space.
 
 The remaining conjectures that have been formulated in \cite{ v05, ov06, o18etal, gk18, oo18etal} 
 are either automatically satisfied in a low-energy approach, or involve unknown parameters that make them less 
 amenable to a quantitive check in an effective field theory context.  The spectrum conjecture, 
 introduced already in \cite{ov06}, is concerned with how the mass spectrum of the theory changes as the field 
 multiplet evolution traverses the field space.  This conjecture is however formulated in an asymptotic form and it contains 
 an unknown parameter, hence there is no model independent estimate of how the spectrum should change in an inflationary 
 theory. Finally, the gradient conjectures, introduced first in \cite{o18etal} and modified in \cite{gk18, oo18etal}, impose 
  constraints on the derivatives of the field theory potential
  that involves a further model dependent parameter whose value has been under
 discussion in the recent literature (see e.g. \cite{a18etal} and references therein). 
 The goal of this conjecture is already achieved if the relevant 
 parameter is strictly positive, if small, in which case compliance of slow-roll inflation with this
 conjecture is immediate.

\vskip .2truein
 
\section{Transfer function dynamics}

In order to evolve the CMB observables given by the spectral indices and the tensor-to-scalar ratios it is useful to 
integrate the dynamics via transfer functions.  Specializing the general adiabatic evolution equation of (\ref{mfi-sr-dynamics}) 
 to twofield inflation gives for conformally flat target spaces with metric factor $G$ 
  \beq
    \cRdot ~=~ 2HG\eta_{\si s} S^{12},
  \eeq
  where $\eta_{\si s}$ is the curved target space slow-roll parameter contraction
  \beq
  \eta_{\si s} ~=~  (\eta_{22}-\eta_{11})\si^1 \si^2 + \eta_{21}(\si^1)^2 - \eta_{12}(\si^2)^2
  \eeq
  that results from the first eq. in (\ref{mfi-sr-dynamics})   in terms of the parameters $\eta_{IJ}$ defined above.
 The isocurvature dynamics specializes to
   \beq
   D_t S^{IJ} 
   ~=~ H\left( 2\e \left(\frac{1}{3}M_\rmPl^2 \cK - 1\right) + \eta_{\si\si} - \eta_{ss}  \right) S^{IJ}, 
   ~~~~~I,J = 1,2,
 \eeq
 where $D_t$ is the covariant time derivative introduced above and $\cK = - 1/\mu^2$ is the Gaussian curvature
 which in the present case directly relates to the Riemann curvature tensor as
 \beq
  R_{IJKL} ~=~ \cK(G_{IK}G_{JL} ~-~ G_{IL}G_{JK}).
  \eeq
 The contraction $\eta_{ss}$ is obtained from the second equation of (\ref{mfi-sr-dynamics}) as 
 \beq
  \eta_{ss} ~=~ \eta_{11}(\si^1)^2 + \eta_{22} (\si^1)^2 - (\eta_{12}+\eta_{21}) \si^1\si^2,
 \eeq
 while $\eta_{\si\si} = \eta_{IJ}\si^I\si^J$.
 
Changing notation for the two-field system by writing $\cS=S^{12}$ these dynamical equations have the general 
form of a coupled system 
\bea
 \cRdot &=& AH \cS \nn \\
 \cSdot &=& B^\G H\cS,
 \eea
 with coefficient functions $A$ and $B^\G$, that can be integrated to lead to
 \beq
  \left(\begin{matrix}  \cR \cr \cS \cr \end{matrix}\right){\Big |}_{t_f}
   ~=~ \left(\begin{matrix} 1 &T_{\cR\cS}\cr 0 &T_{\cS\cS}\cr\end{matrix}\right){\Big |}_{(t_f,t_i)}
         \left(\begin{matrix}  \cR \cr \cS \cr \end{matrix}\right) {\Big |}_{t_i}
\eeq
where the transfer functions take the form
\bea
 T_{\cS\cS}(t) &=&  \exp \left( \int_{t_i}^t dt' ~B^\G(t') H(t')\right) \nn \\
 T_{\cR\cS}(t) &=& \int_{t_i}^t dt'~A(t') H(t') T_{\cS\cS}(t',t_i).
\eea
In the context of a flat field space this was first considered in ref. \cite{w02etal}.

With these transfer functions the evolution of the power spectra $\cP_{\cO\cO'}$  is given by
\bea
 \cP_{\cR\cR}(t) &=& (1+T_{\cR\cS}^2) \cP_{\cR\cR}(t_*) \nn \\
 \cP_{\cR\cS}(t) &=&  T_{\cR\cS} T_{\cS\cS} \cP_{\cR\cR}(t_*) \nn \\
 \cP_{\cS\cS}(t) &=& T_{\cS\cS}^2 \cP_{\cR\cR}(t_*)
\eea
 and the spectral indices evolve as
 \bea
  n_{\cR\cR}(t) &=&  n_{\cR\cR}(t_*) - \left(A_* + B_*^\G T_{\cR\cS}\right) \frac{2T_{\cR\cS}}{1+T_{\cR\cS}^2}\nn \\
  n_{\cR\cS}(t) &=&  n_{\cR\cR}(t_*) - \frac{A_*}{T_{\cR\cS}} - 2B_*^\G  \nn  \\
  n_{\cS\cS}(t) &=& n_{\cR\cR}(t_*) - 2B_*^\G.
\eea
Furthermore the tensor-to-scalar ratio is suppressed by the transfer functions as
 \beq
  r(t) ~=~ \frac{r_*}{1+T_{\cR\cS}^2(t)}.
 \eeq
 
We have seen above that the target space metric $G_{IJ}$ enters in $r_*$, hence the overall structure 
of the metric matters. This is relevant for the most recent of the quantum gravity conjectures, the gradient conjecture.
 In its original formulation this conjecture says that the slow-roll parameter is 
bounded from below by a constant $c$ whose value is not known. If this constant can be argued to be of order one then this 
obviously contradicts slow-roll inflation and the observational bounds of the tensor-to-scalar ratio. For the main purpose
of the conjecture, to ensure the non-existence of de Sitter vacua \cite{dr18}, it is sufficient though to require $c>0$. 
In multifield inflation the natural generalization of the formulation in \cite{o18etal} is to postulate 
$M_\rmPl \sqrt{G^{IJ}V_{,I}V_{,J}} \geq c V$ because this immediately relates to the slow-roll approximation 
 $\e = G^{IJ}\e_I\e_J/2$ of the standard parameter $\e=-\Hdot/H^2$.  
Even within a specific given target space geometry, such as the class of modular inflation 
models, the precise values at horizon crossing $G_{IJ}(\phi_*^I)$ affect $\e_*$, and hence $r_*$, as well as 
the spectral indices, because the metric can vary considerably for different models. 
It is in this context that the structure of the evolution of $r(t)$ becomes of interest because it can in principle 
 enhance or suppress the slow-roll parameters and hence change the CMB parameters at horizon crossing. 
Furthermore, we see  that the transfer functions provide another ingredient that can help suppress the 
physical observable affected by the gradient conjecture, in addition to the effect of the geometry of the target space. 
In general, multifield inflation with more than one isocurvature perturbation $S^{IJ}$ additional transfer functions 
will further suppress the tensor-to-scalar ratio. 

For modular  inflation at higher level the specific form the coefficient functions is somewhat involved.  
The expressions 
\bea
 A &=& 2G\eta_{\si s} \nn \\
 B^\G &=& 2\e \left(\frac{1}{3} M_\rmPl^2 \cK -1\right) + \eta_{\si\si} - \eta_{ss}- \G^I_{IJ}\phidot^J,
\eea
 can be expressed in terms of the basic modular function $F$   because
  the slow-roll parameters are given by \cite{rs16} 
 \bea
  \eta_{\si\si}
    &=&  2 \frac{M_\rmPl^2}{\mu^2}  (\rmIm~\tau)^2 
   \left[ \left|\frac{F'}{F}\right|^2 + \rmRe \left(\frac{F''}{F'}  \frac{\bF'}{\bF}\right)_\rmmod
       - \frac{\pi}{3} \rmIm\left(\Ewhat_2 \frac{\bF'}{\bF}\right)\right]  \nn \\
   \eta_{ss}
  &=&   2 \frac{M_\rmPl^2}{\mu^2}  (\rmIm~\tau)^2 
   \left[ \left|\frac{F'}{F}\right|^2 - \rmRe \left(\frac{F''}{F'}  \frac{\bF'}{\bF}\right)_\rmmod
       + \frac{\pi}{3} \rmIm\left(\Ewhat_2 \frac{\bF'}{\bF}\right)\right] \nn \\
   \eta_{\si s} &=& 
  -2\frac{M_\rmPl^2}{\mu^2} ~(\rmIm~\tau)^2
    \left[ \rmIm\left(\frac{F''}{F'} \frac{\bF'}{\bF}\right)_\rmmod
 ~ + ~ \frac{\pi}{3}~\rmRe\left(\Ewhat_2 \frac{\bF'}{\bF}\right)\right].
\llea{slow-roll-par-general-F}
The specific form of these parameters follows from the results (\ref{Fprime-overF}) and (\ref{Fdoubleprime-overF})
 for the first and second derivatives of the  modular invariant function $F$ at higher level $N>1$.
 These results allow to express the transfer functions in terms of the defining modular forms
$f,g$ and the modular forms $\tf, \tg, \ttf, \ttg, E_{2,N}$, as well as the Eisenstein series $E_2$. Although 
somewhat complicated for the general case their structure simplifies in special cases, such as the model 
of $j$-inflation considered in \cite{rs14,rs16,rs17}, and the models of $h_N$ inflation, to which we turn in the next section.

\vskip .2truein

\section{$h_N$ Inflation}

We have discussed earlier that a natural generalization of the $j$-function and the associated model of $j$-inflation  is given by 
the class of generators of spaces of modular invariant functions associated to Hecke congruence groups $\G_0(N)$.
We define $h_N$-inflation, or $h$-inflation for short, in terms of a specific set of modular functions that belong to the class
 of hauptmodul functions $h_N$, which can be written in terms of the eta-function defined above  as 
  \beq
   h_N(\tau)~:=~ \left(\frac{\eta(\tau)}{\eta(N\tau)}\right)^{24/(N-1)}.
 \lleq{hN}
 Inflaton potentials can be associated to these hauptmodul functions by considering dimensionless functions of 
 the $h_N$, in combination with an energy scale $\Lambda$. A simple set of models is given by powers of the absolute value of $h_N$ as
 \beq
  V_m(\phi^I) ~:=~ \Lambda^4 |h_N(\tau)|^{m},
 \eeq
 where $m$ is an arbitrary integer and the  dimensionless variable $\tau= \tau^1+i\tau^2$ of the complex upper halfplane 
 $\cH$  again parametrizes the inflaton multiplet $\phi^I$ defined in terms of a second energy scale $\mu$ as $\tau^I = \phi^I/\mu$. 
 In the present paper we focus on models with $m=2$:
 \beq
 V(\phi^I) ~:=~ \Lambda^4 |h_N(\tau)|^2.
 \eeq
 
The integers $N$ are required to be such that the integers $(N-1)$ are divisors of $24$, which leads to a class of eight models
which are all associated to genus zero Hecke congruence subgroups, hence the $h_N$ are of hauptmodul type and 
represent generators of the modular invariant functions for the respective
 groups $\G_0(N)$.  This defines the potentials $V_N$ in terms of modular forms $f,g$ of the same weight, 
 given by $w_N=12/(N-1)$,  but different levels $N$. 
  The hauptmodul at $N=2$ in particular is given directly  in terms of the  Ramanujan form $\Delta = \eta^{24}$, 
  which is a cusp form of weight twelve relative to the full modular group 
 \beq
 h_2(\tau) ~=~ \frac{\Delta(\tau)}{\Delta(2\tau)}.
\eeq
 The form $\Delta$ appears in many different contexts because of its fundamental relation to the harmonic oscillator. 
  For higher levels $N>2$ the underlying forms can be thought of as roots of 
 the form $\Delta$ with rescaled arguments.
                              
  Similar to the $j$-function, the hauptmodul functions $h_N$ can be expanded
   as Laurent series whose leading term has order $(-1)$ 
     \beq 
   h_N(q) ~=~ \frac{1}{q} ~+~ \sum_{n\geq 0} a_n^N q^n.
   \eeq
   which again illustrates  that the $h_N$   generalize the  
 $j$-function in a natural way. Writing  $j(\tau) = E_4^3/\eta(\tau)^{24}$ immediately shows that $j$ has the same leading order
 as the $h_p$ functions associated to the higher level groups of genus zero.

 In the following we focus on the subclass associated to  prime $N$ in order to simply the analysis.
 The primes $p$ such that $(p-1)$ divides 24 are the only primes for which $\G_0(p)$ are of genus zero type. This can 
 be seen as follows. The genus of the compact curve $C_p$ obtained from the quotient $\cH/\G_0(p)$ has
 for arbitrary $p$ the genus
 \beq
  g(C_p) ~=~ \frac{p+1}{12} - \frac{\nu_2(p)}{4} - \frac{\nu_3(p)}{3},
  \eeq
  where $\nu_\ell(p)$ is the number of elements $\g$ in $\G_0(p)$ of order $\ell$ such that $\rmtr(\g)<2$.
  Both $\nu_\ell$ are bounded by $\nu_\ell \leq 2$, 
  hence it follows that for $p>13$ the genus of these curves is positive. For the primes $p \leq 13$ the computation 
  of the $\nu_\ell$  shows that for these primes the genus vanishes. 
 
The generators of the groups $\G_0(p)$ include the shift symmetry. This follows from an analysis by
 Rademacher \cite{r29}, who showed that these Hecke congruence subgroups have $2[p/12]+3$ generators, 
 where $p>3$ and $[x]$ denotes the integer part of $x$. Thus for the 
models under consideration here the number of generators is at most five, while for $p=2,3$ there are two generators only. These
generators include the translation $T$ of the full modular group, and hence the shift symmetry.
  
 It is apparent from the definitions (\ref{hN}) and (\ref{dedekind-eta}) that the
  potentials $V_N$ are intricate functions in which the two multiplet components $\phi^I$ interact in a
 highly nonlinear way. The structure of $V_N$ for $N=2,3$ and $N=5$ is illustrated in Fig. 1. This shows that the structure 
 of the $V_N$  potentials  increases in complexity as $N$, leading to an ever more intricate topography.
    \begin{center}
  \includegraphics[scale=0.33]{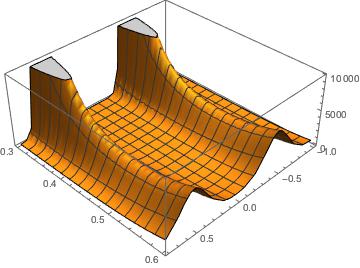} 
  ~~~~~~~~
  \includegraphics[scale=0.33]{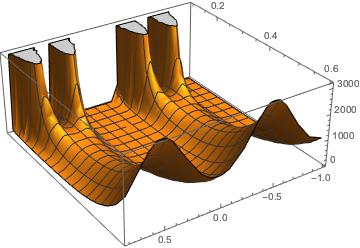} 
  ~~~~~~~
  \includegraphics[scale=0.33]{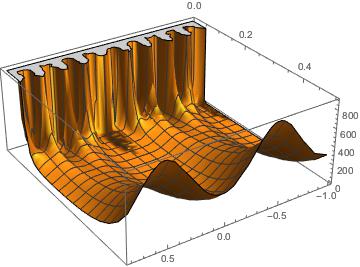}

\begin{quote}
 {\bf Fig. 1}~{\it The structure of the $h_N$ inflation potentials $V/\Lambda^4 = |h_N|^2$ for the modular levels $N=2,3,5$. 
    The plots are not of the same scale. The size of the bounding box is  changed as $N$ increases in order to  make 
    the structure of the individual potentials more transparent.  These graphs illustrate that the topography of the 
    potentials becomes more complex as the level $N$ increases. }
\end{quote}
  \end{center}

These different topographies of the potentials  have an impact on the geometry relevant for  
 the inflationary trajectories and illustrate how different potentials lead to different effects for the same target space metric.
  This in turn is of relevance for the quantum gravity conjectures mentioned in the context of the volume conjecture already.
  Recall that all modular inflation models have the same target space geometry given by the Poincar\'e metric. The different 
  potentials of $h_N$-inflation, as well as $j$-inflation \cite{rs14,rs16,rs17}, however lead to different 
  neighborhoods $U_N(\tau_*^I)$ and $U_j(\tau_*^I)$   of the domain in which the slow-roll parameters are 
  small. Hence the degree of the deviation from the flat limit of the geometry is different for these models as $N$ varies. 
  This affects 
  the gradient conjecture because this immediately translates into a condition on the slow-roll parameter $\e$ and hence 
  the tensor-to-scalar ratio $r$. It turns out that $h_N-$inflation has a stronger metric dependence than $j-$inflation 
  \cite{rs14, rs16}.

\vskip .2truein

 \section{CMB observables for $h_N$-inflation}
 
 We have outlined in the general discussion above that for the phenomenological analysis of the $h$-inflation models 
 it is necessary to evaluate derivatives of the hauptmodul functions. These can be obtained by either specializing our
 results in $\S$4 on higher level inflation to the case $F=h_N$, or they can be computed directly by expressing 
  the Dedekind eta function in terms of Eisenstein  series via the relation of the Ramanujan modular form to the weight 
  four and weight six Eisenstein series as
  \beq
   \Delta ~=~ \frac{E_4^3 - E_6^2}{1728},
 \eeq
 and using Ramanujan's results for the derivatives of $E_w$ \cite{r1916}  (a brief summary can also be found 
 in \cite{rs16}). With these results one obtains
 \beq
 \eta'(N\tau) ~=~ \frac{\pi i}{12} N E_2(N\tau) \eta(N\tau)
 \eeq
 and 
 \beq
  \frac{h'_N}{h_N} ~=~ - \frac{2\pi i}{(N-1)} E_{2,N}.
  \eeq
  This determines the slow-roll parameters $\e_I$ as
  \beq
  \e_I ~=~ - \frac{2\pi i^I}{(N-1)} \frac{M_\rmPl}{\mu} \left(E_{2,N} + (-1)^I \oE_{2,N}\right).
  \eeq
 and leads to the  tensor-to-scalar-ratio as
     \beq
  r ~=~ \frac{128\pi^2}{(N-1)^2} \frac{M_\rmPl^2}{\mu^2} (\rmIm~\tau)^2 |E_{2,N}|^2.
  \lleq{hN-tensor-ratio}
  Alternatively, this expression can be obtained from the general result  above for $r$ for the case of prime levels. 
  For arbitrary $N$ the modular invariance 
  of $r$ arises from the fact that $E_{2,N}$ has weight two. 

The spectral index involves first order and second order derivatives. For hauptmodul inflation the second order derivatives
 can be written in terms of the Eisenstein series $E_w$ as
 \beq
  \frac{h''_N}{h_N} ~=~  \frac{\pi^2}{3(N-1)} \left[ \frac{(N-13)}{(N-1)} E_{2,N}^2 ~-~ E_{4,N} + 2E_{2,N}E_2\right]
  \eeq
  and 
     \beq
    \frac{h''_N}{h'_N} ~=~ \frac{\pi i}{6} \left[\frac{(N-13)}{(N-1)} E_{2,N} - \frac{E_{4,N}}{E_{2,N}}\right]
       ~+~ \frac{\pi i}{3} E_2,
   \eeq
 where we introduce the abbreviation
 \beq
  E_{4,N} ~=~ N^2 E_4^N ~-~ E_4.
 \eeq
 This leads to the adiabatic spectral index
\beq
 n_{\cR\cR}
 = 1 -  \frac{4\pi^2}{3(N-1)} \frac{M_\rmPl^2}{\mu^2} (\rmIm \tau)^2 
        \left[\frac{(N+11)}{(N-1)}  |E_{2,N}|^2  
       -  \rmRe\left( \left( \frac{E_{4,N}}{E_{2,N}} ~-~ 2 \Ewhat_2 \right) \bE_{2,N} \right) 
    \right]
 \lleq{nRR-h-infl}
 where $\Ewhat_2$ is the nonholomorphic modular form defined in (\ref{Ewhat2})
  and $E_{2,N}$ was defined in (\ref{E2levelN}).
 
 In order to determine the number of e-folds it is necessary to integrate the Klein-Gordon equation. In 
 combination with the 
 first Friedman-Lemaitre equation this leads to  the coupled system of equations for $h_N$ inflation 
     given by
 \bea
  \frac{d\tau^1}{ds} ~+~ \frac{4\pi}{\sqrt{3} (N-1)} \frac{\Lambda^2}{\mu^2} \rmIm(\tau)^2 \rmIm(E_{2,N}) ~|h_N| &=& 0 \nn \\
  \frac{d\tau^2}{ds} ~+~ \frac{4\pi}{\sqrt{3} (N-1)} \frac{\Lambda^2}{\mu^2} \rmIm(\tau)^2 \rmRe(E_{2,N}) ~|h_N| &=& 0,
 \llea{kg-h-dynamics}
 where we adopt $s=M_\rmPl t$ as the  dimensionless time variable.
 
Integrating this system allows us to evaluate the number of e-folds $N_k$ via 
 \beq
  N_k ~=~ \frac{\Lambda^2}{\sqrt{3}M_\rmPl} \int dt ~|h_N|
 \lleq{h-inflation-efolds}
 once the energy scale $\Lambda$ has been determined in terms of the amplitude of the power spectrum. 
 The number of e-folds is 
 phenomenologically not precisely constrained and a wide range has been suggested in the literature. In the 
 following we focus on 
 the standard interval, which is given by $N_k\in [50, ~70]$. 
 We call trajectories with $N_k$ in this range as having enough inflation.

The slow-roll parameters (\ref{slow-roll-par-general-F}) specialize to
    \bea
\eta_{\si\si}
  &=& - \frac{2\pi^2}{3(N-1)} \frac{M_\rmPl^2}{\mu^2} (\rmIm~\tau)^2
        \left[ \frac{(N-25)}{(N-1)} |E_{2,N}|^2  
         ~ - ~   \rmRe\left(  \left( \frac{E_{4,N}}{E_{2,N}} - 2\Ewhat_2\right) \oE_{2,N}\right)
     \right] \nn \\
\eta_{ss}
 &=&  \frac{2\pi^2}{3(N-1)} \frac{M_\rmPl^2}{\mu^2} (\rmIm~\tau)^2
        \left[ |E_{2,N}|^2  
         ~ - ~   \rmRe\left(  \left( \frac{E_{4,N}}{E_{2,N}} - 2\Ewhat_2\right) \oE_{2,N}\right)  \right] \nn \\
\eta_{\si s} 
 &=&  - \frac{2\pi^2}{3(N-1)}  \frac{M_\rmPl^2}{\mu^2} (\rmIm~\tau)^2
        \rmIm\left[  \left( \frac{E_{4,N}}{E_{2,N}} - 2\Ewhat_2\right) \oE_{2,N}\right].
 \eea
With these ingredients the evolution of the power spectra, the spectral indices, and the tensor-to-scalar ratio can be 
made explicit in $h_N$-inflation.

\vskip .2truein

\section{Phenomenology of $h_N$ inflation}

For the phenomenological analysis of hauptmodul inflation  we focus on the four main parameters that are commonly used to 
constrain models, the amplitude $A_{\cR\cR}$  and  spectral index $n_{\cR\cR}$ of the adiabatic power spectrum, 
the tensor-to-scalar ratio $r$, and the number of e-folds $N_k$.  These have to be matched with the two energy 
scales $(\Lambda, \mu)$ that define the parameters of the potentials in our framework. The expression  of the 
spectral index and the ratio $r$ depend only on one of the two scales, $\mu$,
which as noted above can be viewed as a proxy for the scalar curvature of the field space.
The  overall scale $\Lambda$ of $h_N$-inflation is determined by 
 the amplitude $\cA_{\cR\cR}$ of the power spectrum via the relation
 \beq
  \cA_{\cR\cR} ~=~ \frac{(N-1)^2}{192\pi^4} \frac{\Lambda^4\mu^2}{M_\rmPl^6} \frac{|h_N|^2}{\rmIm(\tau)^2 |E_{2,N}|^2}.
  \eeq
  While this amplitude does not enter the spectral indices and the gravitational amplitude at horizon crossing, it does 
  determine the evolution of these parameters, hence the number of e-folds.
 The precise values obtained for the observable parameters and their uncertainty in general, and for $\cA_{\cR\cR}$ in particular,   
 depend on the dimensionality of the phenomenological fit and on the precise combination 
 of data sets used in the various analyses \cite{planck18-6, planck18-10, d15etal, d17etal, d18etal}.  Since $\Lambda$ scales 
 as $\cA_{\cR\cR}^{1/4}$ the changes that result from such different choices are slight and do not 
  affect the viability of realization of models located in the central regions of the relevant parameter intervals. 
 Values for $\Lambda$ that are obtained for different trajectories range between $10^{14}\rmGeV$ and $10^{15}\rmGeV$.
 
This leaves the scale $\mu$ as a free parameter. Its range is constrained by the CMB data 
and the number of e-folds $N_k$ that are obtained by solving the dynamical equations (\ref{kg-h-dynamics}) and integrating
(\ref{h-inflation-efolds}).  As noted already, the value of $N_k$ is phenomenologically not well-constrained and a wide range of 
possible values has been discussed in the literature. In our analysis we focus on the standard range $N_k \in [50,~70]$.
 The reported experimental uncertainty of the spectral index is quite small for low-dimensional fits, leading for pure 
 {\sc Planck} data based on TT, TE, EE, lowE and lensing to a value of $0.9649 \pm 0.0042$ in the 
 six-parameter base model.  Adding BAO data changes this to $0.9665 \pm 0.0038$  \cite{planck18-6}.  
 Such low-dimensional fits however do not provide the 
 most  relevant constraints for the landscape of inflationary models in either the singlefield or multifield 
 framework.  In the absence of a multifield statistical analysis with a sufficiently extensive set of parameters we 
 adopt for concreteness 
 a parameter range centered around the value of the {\sc Planck} collaboration 
 $n_{\cR\cR} = 0.96*$. The same uncertainties afflict the reported values for the tensor-to-scalar ratio $r$. 
 The experimental value of this ratio is quite sensitive to what data set is being used and the pure {\sc Planck} data 
 mentioned above for the determination of the spectral index leads to a constraint that is higher than the 
 {\sc Planck}/BICEP2-Keck analysis \cite{a15etal}. The most stringent bounds have been obtained in the final data release 
 of {\sc Planck}  as $r< 0.65$  \cite{planck18-6} and by the BICEP/Keck collaboration as $r< 0.06$ at the pivot scale 
 $k_* = 0.05\rmMpc^{-1}$. These are  weakened if more realistic parameterizations are used. 
 Adopting here the strongest constraint of these collaborations, keeping in mind the unrealistic nature of these fits, 
 we set the upper bound for the tensor ratio $r$ as $r< 0.06$.
 
 The experimental constraints for $n_{\cR\cR}$ and $r$, in  combination with their general multifield expressions
  (\ref{mfi-nRR}) and (\ref{r-ten}),  show that the slow-roll parameters should be small at horizon crossing.
 The $h_N$-inflation  expressions (\ref{nRR-h-infl}) and (\ref{hN-tensor-ratio}) 
 for the spectral index and the tensor ratio $r$ in turn show that the  inflaton values $\tau_*^I$ should 
 be in a region $U(\phi_s^I)$ around a value of the inflaton multiplet for which 
  the modified Eisenstein series $E_{2,N}$ is small.  This suggests scanning the neighborhood $U(\tau_s)$ for 
 $\tau_s = (1+i)/2$. Fig. 2 shows a zoom of this region for the potential at level $N=2$, as well as a few trajectories. 
 It illustrates  that this point is a saddle point of the potential and that the inflaton can roll down 
 from the ridge in either direction, depending on what precisely the value is of $\tau_*$. 
 \begin{center}
 \includegraphics[scale=0.4]{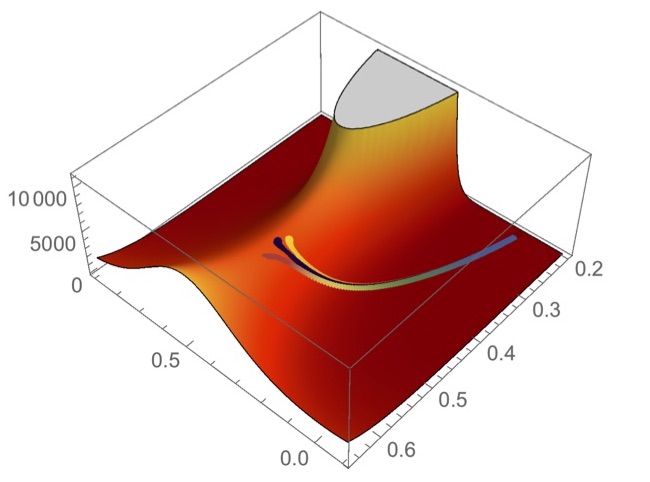}
  \end{center}
 \begin{quote}
 {\bf Fig. 2.}~{\it A zoom of the neighborhood $U(\tau_s)$ of the potential surface of $V/\Lambda^4=|h_2|^2$, where 
    $\tau_s$ is the zero of the function $E_{2,2}$.
      The curves on the surface illustrate a selection of inflationary trajectories that are compatible with the 
    observational CMB constraints obtained in the recent literature \cite{planck18-6, bk18}.}
 \end{quote}
 
 The detailed form of the inflationary path depends on whether the initial value is closer or farther from the open boundary 
 of target space given by the upper halfplane given by the real axis. Depending on whether this point is on the near or the 
 far side of the saddle point 
 the trajectory will make  a turn at the beginning of its evolution. This becomes more transparent in Figure 2, 
 which shows trajectories with sufficient inflation and parameters compatible with the CMB constraint of
  the {\sc Planck} and BICEP collaborations.  Our numerical analysis of the inflaton target space for 
 $h_2$ inflation with a step size of $\Delta \tau^I = 10^{-3}$ produced values  for the tensor ratio $r$ values 
 in the range
  \beq
   0.0019 ~\leq ~ r ~ \leq ~0.06.
  \eeq
  These values provide a target for experiments that are currently under constructions, such as 
     the BICEP array and SPT-3G  at the South Pole \cite{bk18, spt3g14},  
     the CLASS and 
     the Simons observatory in the Atacama desert \cite{class18, so18}, and 
     the Ali-CPT in Tibet \cite{ali-cpt17},
  among others. The CLASS experiment aims to reach $r=0.01$, while the other experiments aim to reach
   target sensitivities $\si(r) = 10^{-3}$ within the next few years,  allowing them to 
  nontrivially constrain the parameter space of $h$-inflation.  In Figure 3 we place several of the $h_2$-models 
   in the $(n_{\cR\cR}, r)$ plane as determined by the {\sc Planck} Collaboration 
    \cite{planck18-6}. In this plot the individual trajectories are obtained by fixing the initial value of the dimensionless
     inflaton and letting the energy scale $\mu$ run 
   within an interval of $\Delta \mu/M_\rmPl = \pm 2$ of the value identified by the phenomenological constraints, 
   as described above.
   
   \begin{center}
   \includegraphics[scale=0.07]{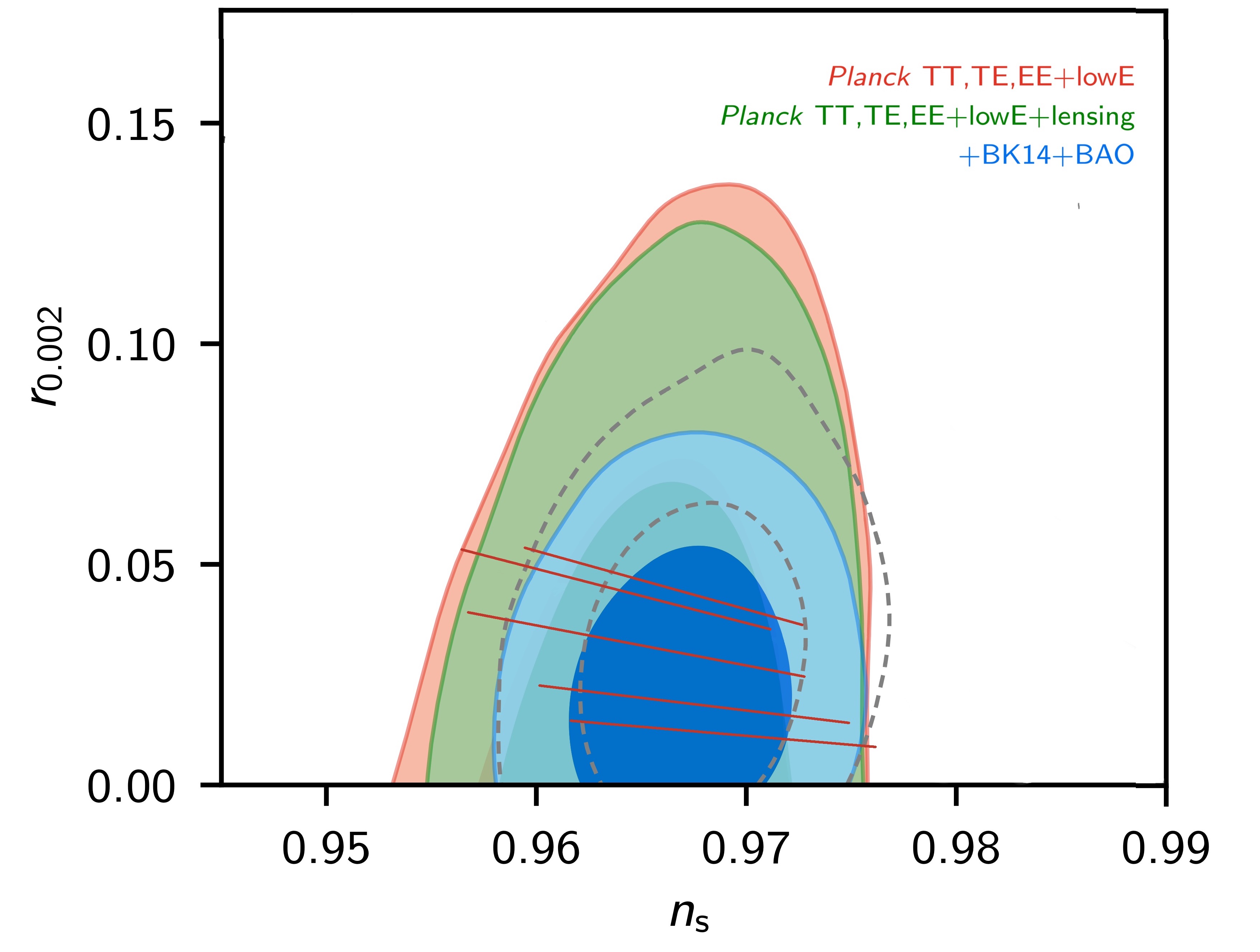}
   \end{center}
   \begin{quote}
   {\bf Fig. 3}~ {\it An illustration of the tensor-to-scalar ratio $r$ for $h_2$-inflation. The lines are 
     $\mu$-trajectories in the plane spanned by the spectral index and $r$ as obtained in
   the final {\sc Planck} data release. The detailed description of the various data sets can be 
   found in \cite{planck18-6}.}
   \end{quote}
   
   \vskip .3truein
   
 In the discussion above of the general structure of modular inflation we have noted that the Poincar\'e metric is common to all
  such models, hence the geometric distortion determined by the metric in field space has a universal 
 effect on the trajectories as they evolve. Since the magnitude of the metric coefficients increases as the trajectory approaches 
 the open boundary of the field space, the impact of the nontrivial geometry becomes more pronounced close to this 
 boundary.  While this effect is universal, the details of this impact of the  geometry will nevertheless be
  different for different potentials because the specific regions in the 
 target space that give rise to enough inflation depend on the potential. In the case of $h$-inflation 
  it turns out that it is the later part of the trajectory that is more affected by the metric than the 
 earlier evolution. This is illustrated in Fig. 4 via an intensity map. 
 
 \begin{center}
\includegraphics[scale=0.3]{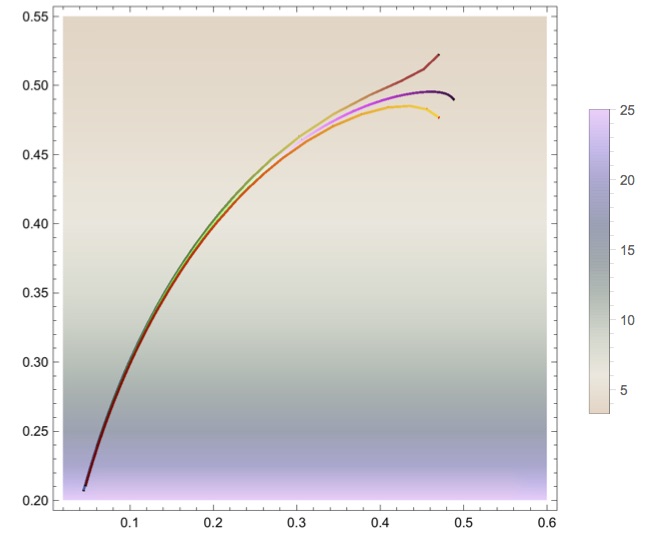}
 \end{center}
 \begin{quote}
 {\bf Fig. 4.}~{\it An illustration of phenomenologically consistent trajectories for the inflationary 
           model based on $h_2$.
    The intensity colors indicate the strength of the Poincar\'e metric, which increases as the imaginary component of the 
    complex inflaton $\tau$ decreases.}
    
 \end{quote}
 
 For $N>2$ a similar analysis can be done. As shown in Fig. 1, the ridge structure of the potential surface becomes
  more intricate for 
 higher $N$.  The zeros of the functions $E_{2,N}$ that enter the slow-roll parameter $\e$, and hence the CMB observables, 
 vary with $N$, hence the neighborhoods $U(\tau_*)$ are located in different regions of the target space. 
 This also implies that the metric
values will be different for these different higher level models, and hence the geometric effects on the observables will 
vary with  $N$. 

 A more detailed analysis of the constraints implied by the satellite experiments necessitates the adoption of specific scenarios 
 that parametrize the link between the field theoretic isocurvature perturbation and the isocurvature perturbations
  that are constrained by WMAP \cite{h12etal} and the {\sc Planck} collaboration
   \cite{planck13, planck15, planck18-6, planck18-10}. 
  The {\sc Planck} group in particular has provided 
  constraints for cold dark matter, neutrino density and neutrino velocity isocurvature perturbations, collectively denoted by 
   $\cI_a$ in \cite{planck15, planck18-6, planck18-10}.  These can be used to 
 provide bounds on the corresponding fractions  $U_a$ induced by the twofield isocurvature. 
 A further extension of the work presented here can also include the period of reheating by generalizing the 
 strategy developed in \cite{rs17} for $j$-inflation to higher levels. This analysis will be left for future work.

\vskip .2truein

\section{Conclusion}

In this paper we have introduced modular inflation at higher level $N>1$, which are theories with 
 an underlying symmetry group $\G_N$ embedded in the modular group $\rmSL(2,\mathZ)$. 
  For concreteness we have restricted our focus on the particular class given by Hecke congruence 
subgroups $\G_N=\G_0(N)$, because for these groups the theory of modular forms is most developed. 
This generalizes the framework introduced in ref. \cite{rs16}, where the discussion  is restricted to level one. 
One of the key issues encountered at higher level is the modular nature of the observables because 
the phenomenological analysis introduces derivatives of modular forms that are not themselves modular. We 
have shown that the modularity structure found in \cite{rs16} for the full modular group generalizes to higher level, 
although the modular building blocks are very different when $N>1$.
As in the case of level one, the spectral indices as well as the tensor-to-scalar ratio are  modular in a generalized 
sense in that modularity is more important than the holomorphic structure that the theory starts out with. 
While the holomorphic 
part of the observables  is completely different from the structure at level one, it  turns out that again the 
nonholomorphic modular form $\Ewhat_2$ is sufficient to ensure the modularity  of the 
phenomenological variables.

In order to make the framework of modular inflation at higher level $N>1$ 
concrete we have also introduced the class of $h_N$-inflation models based on 
hauptmodul functions, or principal modul functions. As in the case of $j$-inflation these functions are distinguished by the 
fact that they are generators of the space of inflation potentials at a given level. 
In the general discussion of the paper we have made the structure of modular inflation 
explicit for these theories. 
We have focused on the special case of prime levels because the analysis simplifies considerably, even though it remains
more involved than at level one.
It would be of interest to consider the explicit structure of the models introduced here at nonprime levels.
For the specific $h_N$ model at $N=2$ we have analyzed the CMB observables $n_{\cR\cR}, r, N_k$ and shown that 
in particular the tensor ratio $r$ provides target values that are accessible to ground-based experiments that are currently under 
construction and that are scheduled to reach their design sensitivity within the next few years.

In the course of the development of the general framework of modular inflation at higher level $N$, we have also 
 considered features of these theories that are relevant for  issues related to the possible UV completion of 
these models.  In the context of this discussion a number of conjectures have been formulated under the heading of 
the swampland, introduced by Vafa, that pose restrictions on
the topology and geometry of the target space and that also constrain the types of potentials. The status of these
constraints remains under vigorous discussion. We have shown here 
that those conjectures that admit a precise formulation are satisfied in modular inflation while most of the 
multifield inflation theories considered in the literature are in conflict with the geometric constraints and hence 
belong to the swampland. Since the same ingredients that make modular inflation compatible with the quantum 
gravity conjectures operate 
 in general automorphic inflation \cite{rs14, rs15} this more general multifield inflation framework will be in compliance with 
 the constraints as well.

\vskip .2truein

{\large {\bf Acknowledgement.}} \\
This work was supported in part by a Faculty Research Grant of Indiana University South Bend.

\vskip .3truein

\baselineskip=17pt
\parskip=0.02pt

{\large {\bf References}}

\end{document}